\documentclass[a4paper,11pt]{article}

\usepackage{jcappub} 
\usepackage{hyperref}
\usepackage[T1]{fontenc} 
\usepackage{multirow,subcaption}

\title{\boldmath A Unified Explanation of Gamma-Ray and Neutrino Spectra from Astrophysical Sources Based on the Gluon Condensation Model}

\author{Jiangyuan Qian,}
\author{Jintao Wu,}
\author[1]{Jianhong Ruan\note{Corresponding author.}}

\affiliation{Department of Physics, East China Normal University,\\Shanghai 201100, China}

\emailAdd{3492568051@qq.com}
\emailAdd{jin-taowu@qq.com}
\emailAdd{jhruan@phy.ecnu.edu.cn}

\abstract{The advent of multi-messenger astronomy has provided abundant information for understanding the acceleration and particle-production mechanisms of cosmic rays. In this work, we present a unified study of cosmic gamma-ray and neutrino spectra within the Gluon Condensation (GC) model. Derived from Quantum Chromodynamics (QCD), the GC model predicts that, in high-energy hadronic processes, gluons may condense near a critical momentum, leading to a dramatic enhancement in secondary-pion production and imprinting a characteristic broken power-law feature on the gamma-ray spectrum. Within this framework, we first derive the neutrino spectrum corresponding to the GC scenario and then investigate three astrophysical sources with both gamma-ray observations and neutrino candidate signals: the active galactic nuclei TXS~0506+056 and NGC~1068, and the supernova remnant G54.1+0.3. Using the GC model, we fit the observed gamma-ray spectra of these sources and predict their corresponding neutrino spectra. Our results show that the gamma-ray spectra of TXS~0506+056 and NGC~1068 are well described by the GC model, and that the predicted neutrino spectra are consistent with IceCube observations within uncertainties; in particular, clear relations are found between their relative magnitudes. For SNR~G54.1+0.3, however, the GC-predicted neutrino spectrum exhibits continuous hardening after the break, deviating from the typical power-law behavior expected for cosmic-ray secondaries and thus disfavoring a common GC origin. This study represents the first systematic attempt to correlate gamma-ray and neutrino spectra within the GC framework, offering a new perspective on multi-messenger emission from high-energy astrophysical sources.}
	

\begin{document}
\maketitle
\flushbottom

\section{Introduction}
\label{sec:intro}
The origin and acceleration mechanisms of high-energy cosmic rays (CRs) remain among the most fundamental challenges in modern astrophysics. In violent astrophysical environments such as supernova remnants (SNRs), gamma-ray bursts (GRBs), and active galactic nuclei (AGN), high-energy collisions—such as proton-proton (\(pp\)) and proton-photon (\(p\gamma\)) interactions—produce copious secondary particles, including high-energy gamma-rays and neutrinos \cite{1,2}. In recent decades, significant progress in multi-messenger astronomy has been driven by ground-based and space-borne observatories. Gamma-ray detectors such as Fermi-LAT, VERITAS, MAGIC, and LHAASO have covered energies from keV to PeV, while the IceCube Neutrino Observatory has accumulated over a decade of data, identifying candidate neutrino sources such as the blazar TXS~0506+056  \cite{3} and the Seyfert galaxy NGC~1068 \cite{4,5}. These AGN have become key targets for understanding hadronic acceleration processes. More recently, the supernova remnant G54.1+0.3 has also been suggested as a potential neutrino emitter \cite{6}, further expanding the landscape of candidate multi-messenger sources.

The limited number of astrophysical sources with coincident gamma-ray and neutrino detections makes them particularly valuable for probing the underlying particle acceleration and interaction mechanisms. 
The simultaneous production of gamma-rays and neutrinos in hadronic collisions has long been recognized as a key signature of cosmic-ray acceleration in astrophysical sources. Previous studies have explored the connection between these two messengers within various theoretical frameworks, including one-zone lepto-hadronic models \cite{7,8}, two-zone scenarios involving distinct emission regions \cite{9,10}, and phenomenological scaling relations between gamma-ray and neutrino luminosities \cite{11,12}. These approaches have provided valuable constraints on the physical conditions of the emission regions and the efficiency of hadronic processes. However, a unified theoretical framework that can simultaneously describe the spectral shapes of both gamma-rays and neutrinos from individual sources, while being grounded in first-principles QCD, has remained elusive.

In recent years, the gluon condensation (GC) model has emerged as a novel theoretical framework grounded in quantum chromodynamics (QCD). At sufficiently high collision energies, gluons inside protons are predicted to condense near a critical momentum, triggering strong shadowing and anti-shadowing effects that dramatically enhance the production of secondary pions \cite{13,14,15}. This GC effect imprints a characteristic broken power-law feature on the gamma-ray spectrum and has been successfully applied to explain the gamma-ray observations of the Tycho supernova remnant \cite{16} and several AGNs \cite{17}. Moreover, the GC model naturally connects the spectra of various secondary particles, including gamma-rays, electrons, and protons, that originate from the same hadronic process \cite{18,19,20}. However, the relationship between gamma-ray and neutrino spectra within the GC framework has not yet been systematically explored. In particular, the GC model provides explicit predictions for the spectral indices of both gamma-rays (from \(\pi^0\) decay) and neutrinos (from \(\pi^\pm\) decay) produced in the same hadronic process. This predictive power distinguishes the GC model from purely phenomenological approaches and allows quantitative tests against multi-messenger observations.

In this work, we aim to investigate the possible connection between gamma-ray and neutrino emissions from astrophysical sources within the GC model. For sources with coincident gamma-ray and neutrino detections, if the observed gamma-ray spectrum exhibits GC characteristics, the corresponding neutrino spectrum can be predicted. We select three candidate sources—TXS~0506+056, NGC~1068, and SNR~G54.1+0.3—whose gamma-ray spectra show clear broken power-law features and for which neutrino signals have been reported. Using the GC model, we fit their gamma-ray spectra and derive the expected neutrino spectra. Our results show that the gamma-ray spectra of TXS~0506+056 and NGC~1068 are well described by the GC model, and the predicted neutrino spectra are consistent with IceCube observations within uncertainties, suggesting a common GC origin. In contrast, the predicted neutrino spectrum for SNR~G54.1+0.3 exhibits an unphysical hardening after the break, disfavoring a GC interpretation. This study represents the first systematic attempt to correlate gamma-ray and neutrino spectra within the GC framework, offering a new perspective on the multi-messenger emission from high-energy astrophysical sources.

The remainder of this paper is organized as follows. Section~\ref{sec:2} introduces the GC model and derives the expressions for the GC spectra of gamma-rays and neutrinos. Section~\ref{sec:3} presents the analysis of the three sources: TXS~0506+056, NGC~1068, and SNR~G54.1+0.3. Section~\ref{sec:4} summarizes the main conclusions and discusses future prospects.

\section{The GC Model}
\label{sec:2}
Secondary cosmic-ray particles may originate from hadronic processes such as $p+p(A)\to \pi^\pm + \pi^0 + p+\overline{p}+ \text{other}$, followed by subsequent decays such as $\pi^0\to 2\gamma$ and the decay chain of $\pi^\pm$. We define $N_\pi(E_{p-p(A)},E_\pi)$ as the number of mesons with energy $E_\pi$ produced in $p-p(A)$ collisions, where $E_{p-p(A)}$ is the energy of the incident proton in the rest frame of the target proton. Owing to the non-perturbative hadronization process in $p-p(A)$ collisions, the distribution of $\pi$ mesons is difficult to calculate directly. To simplify the analysis, the GC model considers mesons as the dominant secondary particles, since their abundance is much greater than that of other species in high-energy collisions. Typically, these $\pi$ mesons carry relatively low kinetic energy (or momentum) in the center-of-mass (CM) frame, especially in the central rapidity region. At a given interaction energy, the maximum value of $N_\pi$ can occur only when almost all available kinetic energy in the CM frame is converted into $\pi$ production. We therefore assume that a large number of gluons are produced in the central region due to the GC effect, leading to the maximum value of $N_\pi$. It should be emphasized that this assumption is introduced only to simplify the calculation and is not itself a requirement of the GC condition. As shown below, it greatly simplifies the derivation without fundamentally altering the characteristic features of GC. By considering relativistic invariants and energy conservation, one obtains the following equation (\cite{21}):
\begin{equation}\label{eq:2.1}
	\begin{aligned}
		(2m_p^2+2E_{p-p(A)}m_p)^{1/2} &= E_{p1}^* + E_{p2}^* + N_{\pi} m_{\pi},\\
		E_{p-p(A)} + m_p &= m_p \gamma_1 + m_p \gamma_2 + N_{\pi} m_{\pi} \gamma.
	\end{aligned}
\end{equation}
Here, $E_{(p_i)}^*$ is the energy of the leading protons in the CM system, $\gamma_i$ is the Lorentz factor for the corresponding particles. Considering the parameter $K \sim 0.5$ \cite{22}, the eq.~(\ref{eq:2.1}) can be simplified to:
\begin{equation}\label{eq:2.2}
	\begin{aligned}
		E_{p1}^* + E_{p2}^* &= \left(\frac{1}{K} - 1\right) N_{\pi} m_{\pi},\\ 
		m_p \gamma_1 + m_p \gamma_2 &= \left(\frac{1}{K} - 1\right) N_{\pi} m_{\pi} \gamma.
	\end{aligned}
\end{equation}
For $p-p(A)$ collisions, the solution for $N_{\pi}(E_{p-p(A)}, E_{\pi})$ can be expressed as:
\begin{equation}\label{eq:2.3}
	\begin{aligned}
		\ln N_{\pi} &= 0.5 \ln E_{p-p(A)} + a, \\
		\ln N_{\pi} &= \ln E_{\pi} + b,
	\end{aligned}
\end{equation}
where
\begin{equation}\label{eq:2.4}
	\begin{aligned}
		a &\equiv 0.5 \ln (2m_p) - \ln m_{\pi} + \ln K, \\
		b &\equiv \ln (2m_p) - 2 \ln m_{\pi} + \ln K.
	\end{aligned}
\end{equation}
In these equations, $E_{\pi} \in [E_{\pi}^{\mathrm{GC}}, E_{\pi}^{\text{max}}]$. eq.~(\ref{eq:2.3}) establishes a direct relationship among $N_{\pi}$, $E_{p-p(A)}$, and $E_{\pi}^{\mathrm{GC}}$, thereby facilitating the derivation of the gamma-ray spectrum in the GC model.

In general, the gamma-ray spectrum is given by:
\begin{equation}\label{eq:2.5}
	\Phi_{\gamma}(E_{\gamma}) = \Phi_{\gamma}^0(E_{\gamma}) + \Phi_{\gamma}^{GC}(E_{\gamma}).
\end{equation}
Here, $\Phi_{\gamma}^0(E_{\gamma})$ denotes the background gamma-ray spectrum. In the GC model, the local gamma-ray contribution is given by:
\begin{equation}\label{eq:2.6}
	\begin{aligned}
		\Phi_{\gamma}^{GC}(E_{\gamma}) &= C_{\gamma} \left(\frac{E_{\gamma}}{E_0}\right)^{-\beta_{\gamma}} \int_{E_{\pi}^{\text{min}}}^{E_{\pi}^{\text{max}}} \mathrm{d}E_{\pi} \left(\frac{E_{p-p(A)}}{E_{p-p(A)}^{GC}}\right)^{-\beta_p}\\
		&\times N_{\pi}(E_{p-p(A)}, E_{\pi}) \left(\frac{\mathrm{d}\omega_{\pi-\gamma}(E_{\pi},E_{\gamma})}{\mathrm{d}E_{\gamma}}\right).
	\end{aligned}
\end{equation}
In this equation, $\beta_\gamma$ characterizes the propagation losses of gamma-rays, and $\beta_p$ is related to the proton acceleration mechanism. The parameter $C_{\gamma}$ combines the kinematic factors and the dimensionality of the proton spectrum with the branching ratio of the $\pi^0 \to 2\gamma$ process. The normalized spectrum for $\pi^0 \to 2\gamma$ is given by:
\begin{equation}\label{eq:2.7}
	\frac{\mathrm{d}\omega_{\pi-\gamma}(E_{\pi}, E_{\gamma})}{\mathrm{d}E_{\gamma}} = \frac{2}{\beta_{\pi} E_{\pi}} H\left[E_{\gamma}; \frac{1}{2} E_{\pi} (1 - \beta_{\pi}), \frac{1}{2} E_{\pi} (1 + \beta_{\pi})\right],
\end{equation}
where $H(x;a,b)$ represents the Heaviside function, which equals 1 when $a \leq x \leq b$ and 0 otherwise. By substituting Eqs.~\ref{eq:2.3}, \ref{eq:2.4}, and \ref{eq:2.7} into eq.~(\ref{eq:2.6}), the GC-characteristic gamma-ray spectrum is obtained:
\begin{equation}\label{eq:2.8}
	\begin{aligned}
		\Phi_{\gamma}^{GC}(E_{\gamma}) &= C_{\gamma} \left(\frac{E_{\gamma}}{E_{\pi}^{\text{GC}}}\right)^{-\beta_{\gamma}} \int_{E_{\pi}^{\text{GC}} \text{ or } E_{\gamma}}^{E_{\pi}^{\text{GC,max}}} \mathrm{d}E_{\pi} \left(\frac{E_{p-p(A)}}{E_{p-p(A)}^{\text{GC}}}\right)^{-\beta_p} \\
		&\times N_{\pi}(E_{p-p(A)}, E_{\pi}) \frac{2}{\beta_{\pi} E_{\pi}}.
	\end{aligned}
\end{equation}
If $E_\gamma \leq E_{\pi}^\mathrm{GC}$ (or $E_\gamma>E_{\pi}^\mathrm{GC}$), the lower limit of the integral is $E_\pi^\mathrm{GC}$ (or $E_\gamma$). Finally, integrating this expression gives the parameterized form of gamma-ray \cite{17,24}:
\begin{equation}\label{eq:2.9}
	\Phi_{\gamma}^{\mathrm{GC}}(E_{\gamma}) =
	\begin{cases} \displaystyle
		\frac{50C_\gamma}{2\beta_p - 1} E_{\pi}^{\mathrm{GC}} \left(\frac{E_{\gamma}}{E_{\pi}^{\mathrm{GC}}}\right)^{-\beta_{\gamma}} & \text{if } E_{\gamma} \leq E_{\pi}^{\mathrm{GC}}, \\ \displaystyle
		\frac{50C_\gamma}{2\beta_p - 1} E_{\pi}^{\mathrm{GC}} \left(\frac{E_{\gamma}}{E_{\pi}^{\mathrm{GC}}}\right)^{-\beta_{\gamma} - 2\beta_p + 1} & \text{if } E_{\gamma} > E_{\pi}^{\mathrm{GC}}.
	\end{cases}
\end{equation}
This parameterized form is a broken power law with a sharp spectral break. The pure power-law behavior for $E_{\gamma} \leq E_{\pi}^{\mathrm{GC}}$ arises from the fixed lower limit of integration in eq.~(\ref{eq:2.8}), namely $E_\pi^\mathrm{GC}$. These features of $\Phi_\gamma^\mathrm{GC}$ are directly induced by the GC effect and are referred to as GC features. They differ from the smooth spectral shapes usually expected from other radiation mechanisms. In eq.~(\ref{eq:2.9}), the second power-law branch for $E_\gamma>E_\pi^\mathrm{GC}$ results from the simplifying assumption in eq.~(\ref{eq:2.1}), namely that all available kinetic energy in the central region is used to produce $\pi$ mesons. If this simplification is relaxed, the behavior of the spectrum for $E_\gamma>E_\pi^\mathrm{GC}$ under the integral conditions may vary. This parameterized form can therefore be compared with experimentall data to test the validity of the simplification, while the essential GC features remain unchanged.

Considering the processes $\pi^{\pm} \rightarrow \mu^{\pm} + \nu_{\mu} (\bar{\nu}_{\mu})$ and $\mu^{\pm} \rightarrow e^{\pm} + \nu_e (\bar{\nu}_e) + \bar{\nu}_{\mu} (\nu_{\mu})$, which produce neutrinos and electrons, the GC neutrino spectrum can be derived as:
\begin{equation}\label{eq:2.10}
	\begin{aligned}
		\Phi_{\nu}^{\mathrm{GC}}(E_{\nu}) &= C_{\nu} \left(\frac{E_{\nu}}{E_{\pi}^{\mathrm{GC}}}\right)^{-\beta_{\nu}} \int \mathrm{d}E_{\mu} \int_{E_{\pi}^{\mathrm{GC}} \text{ or } E_{\nu}}^{E_{\pi}^{max}} \frac{\mathrm{d}E_{\pi}}{E_{\pi}} \left(\frac{E_{p-p(A)}}{E_{p-p(A)}^{\mathrm{GC}}}\right)^{-\beta_p}\\ 
		&\times N_{\pi^{\pm}}(E_{p-p(A)}, E_{\pi}) \left(\frac{\mathrm{d}\omega_{\pi-\mu}(E_{\pi},E_{\mu})}{\mathrm{d}E_{\mu}}\right) \left(1 + \frac{\mathrm{d}\omega_{\mu-\nu}(E_{\mu},E_{\nu})}{\mathrm{d}E_{\nu}}\right)\\
		&= C_{\nu} E_{\pi}^{\mathrm{GC}}\left(\frac{E_{\nu}}{E_{\pi}^{\mathrm{GC}}}\right)^{-\beta_{\nu}} \\
		&\times 
	\begin{cases}
		\left(\frac{-31}{2\beta_p - 1} + \frac{240}{\beta_p}\right) \left(\frac{E_{\nu}}{E_{\pi}^{\mathrm{GC}}}\right) & \text{if } E_{\nu} \leq E_{\pi}^{\mathrm{GC}}, \\ 
	   \left(\frac{-31}{2\beta_p - 1} + \frac{240}{\beta_p} \right) \left(\frac{E_{\nu}}{E_{\pi}^{\mathrm{GC}}}\right)^{-2\beta_p + 1} & \text{if } E_{\nu} > E_{\pi}^{\mathrm{GC}}.
	\end{cases}
	\end{aligned}
\end{equation}
The normalized spectra appearing in the above integration are given by:
\begin{equation}\label{eq:2.11}
	\begin{aligned}
		\frac{\mathrm{d}\omega_{\pi-\mu}(E_{\pi}, E_{\mu})}{\mathrm{d}E_{\mu}} &= \delta(E_{\mu} - 0.8E_{\pi}),\\
		\frac{\mathrm{d}\omega_{\mu-\nu}(E_{\mu}, E_{\nu})}{\mathrm{d}E_{\nu}} &= 16\left(1 - \frac{E_{\nu}}{E_{\mu}}\right)^2 \left(\frac{2E_{\nu}}{E_{\mu}} - 0.5\right).
	\end{aligned}
\end{equation}
It is important to note that the lower limit of the integral in eq.~(\ref{eq:2.10}) is $\max(E_{\pi}^{\mathrm{GC}}, E_{\nu})$, which also leads to a sharp break around $E_{\nu} = E_{\pi}^{\mathrm{GC}}$.

As can be seen from eq.~(\ref{eq:2.10}), the neutrino spectrum is governed by the parameters $\beta_p$, $\beta_\nu$, $E_{\pi}^{\mathrm{GC}}$, and $C_\nu$. Provided that the neutrinos and gamma-rays share a common GC origin, the parameters $\beta_p$ and $E_{\pi}^{\mathrm{GC}}$ can be determined from the gamma-ray spectrum. Under the assumption of negligible attenuation during neutrino propagation, i.e., by setting $\beta_\nu = 0$, the neutrino spectrum is reduced to a single free parameter, $C_\nu$, which may be constrained by experimentall data. Consequently, within the GC framework, the neutrino spectrum can be predicted from the gamma-ray spectrum.

It is worth noting that when $E_{\gamma}=E_{\nu}=E_{\pi}^{\mathrm{GC}}$, dividing eq.~(\ref{eq:2.9}) by eq.~(\ref{eq:2.10}) yields:
\begin{equation}\label{eq:2.12}
	\begin{aligned}
	R_{\gamma \nu}=	\frac{\Phi_{\gamma}^{\mathrm{GC}}}{\Phi_{\nu}^{\mathrm{GC}}} &= \frac{C_{\gamma}}{C_{\nu}}\left(\frac{\frac{50}{2\beta_p - 1}}{{\frac{-31}{2\beta_p - 1} +\frac{240}{\beta_p}}}\right)=\frac{C_{\gamma}}{C_{\nu}}\frac{\beta_p}{8.98\beta_p-4.8}.
	\end{aligned}
\end{equation}
Since the parameters $C_{\gamma}$ and $C_{\nu}$ depend on the kinematic factors and the correponding decay branching ratios, in principle the ratio $C_{\gamma}/C_{\nu}$ should take a common value for different sources within the same framework. Therefore, if we consider two distinct sources and compare their $R_{\gamma \nu}$ values, we obtain

 \begin{equation}\label{eq:2.13}
 	\begin{aligned}
 		\frac{R_{1\gamma\nu}}{R_{2\gamma\nu}}= \frac{\beta_{p1}}{\beta_{p2}}\frac{8.98\beta_{p2}-4.8}{8.98\beta_{p1}-4.8}.
 	\end{aligned}
 \end{equation}
This relation provides a way to connect two different sources. In the following analysis of TXS~0506+056 and NGC~1068, we find that their spectra are indeed broadly consistent with this relation.

\section{Analysis of Astrophysical Sources}
\label{sec:3}
In this section, we apply the GC model to three sources with both gamma-ray and neutrino observations: TXS~0506+056, NGC~1068, and SNR~G54.1+0.3. For each source, we fit the observed gamma-ray spectrum using eq.~(\ref{eq:2.9}) to determine the relevant parameters, then predict the corresponding neutrino spectrum using eq.~(\ref{eq:2.10}) and compare it with IceCube observations.

\subsection{Source TXS 0506+056}
\label{sec:3.1}

TXS 0506+056 is a blazar at redshift \(z = 0.3365 \pm 0.0010\) and one of the brightest gamma-ray sources in the sky. It is the first astrophysical object—apart from the Sun and Supernova 1987A—confirmed to have a high probability of emitting neutrinos. On September 22, 2017, the IceCube detector recorded a 290~TeV neutrino event (IceCube-170922A) spatially and temporally coincident with this source, with a reconstructed arrival direction within \(0.1^\circ\) of TXS~0506+056 and a statistical significance exceeding \(3.5\sigma\) \cite{25,26}. The neutrino energy is estimated to be 290~TeV, with a lower limit of 183~TeV and an upper limit of 4.3~PeV. Optical observations further revealed that the source transitioned from a “closed” to an “open” state two hours after the neutrino emission, providing strong temporal evidence for the association \cite{26}.Later, an independent search for point-like sources in the northern hemisphere using ten years of IceCube data revealed that TXS 0506+056 is coincident with the second hottest-spot of the neutrino event excess \cite{5}. 

The physical origin of the neutrinos in TXS~0506+056 remains debated. Two main scenarios have been proposed: production in the relativistic jet, or production in the corona near the supermassive black hole. In the jet scenario, protons accelerated in the jet interact with ambient photons or matter via \(p\gamma\) or \(pp\) interactions, producing neutrinos along with gamma-rays that can escape with minimal attenuation \cite{8,10,28,29,7,32,9,34,35,36,38,39}. In the coronal scenario, the dense X-ray field in the corona renders the region opaque to high-energy gamma-rays, which are reprocessed into MeV photons via electromagnetic cascades, while neutrinos escape freely \cite{40,41,42}. In this work, we only consider the jet scenario, since in the corona scenario, we cannot directly obtain the original gamma-ray spectrum.

\begin{figure}[htbp]
	\centering 
	{\includegraphics[width=0.9\textwidth]{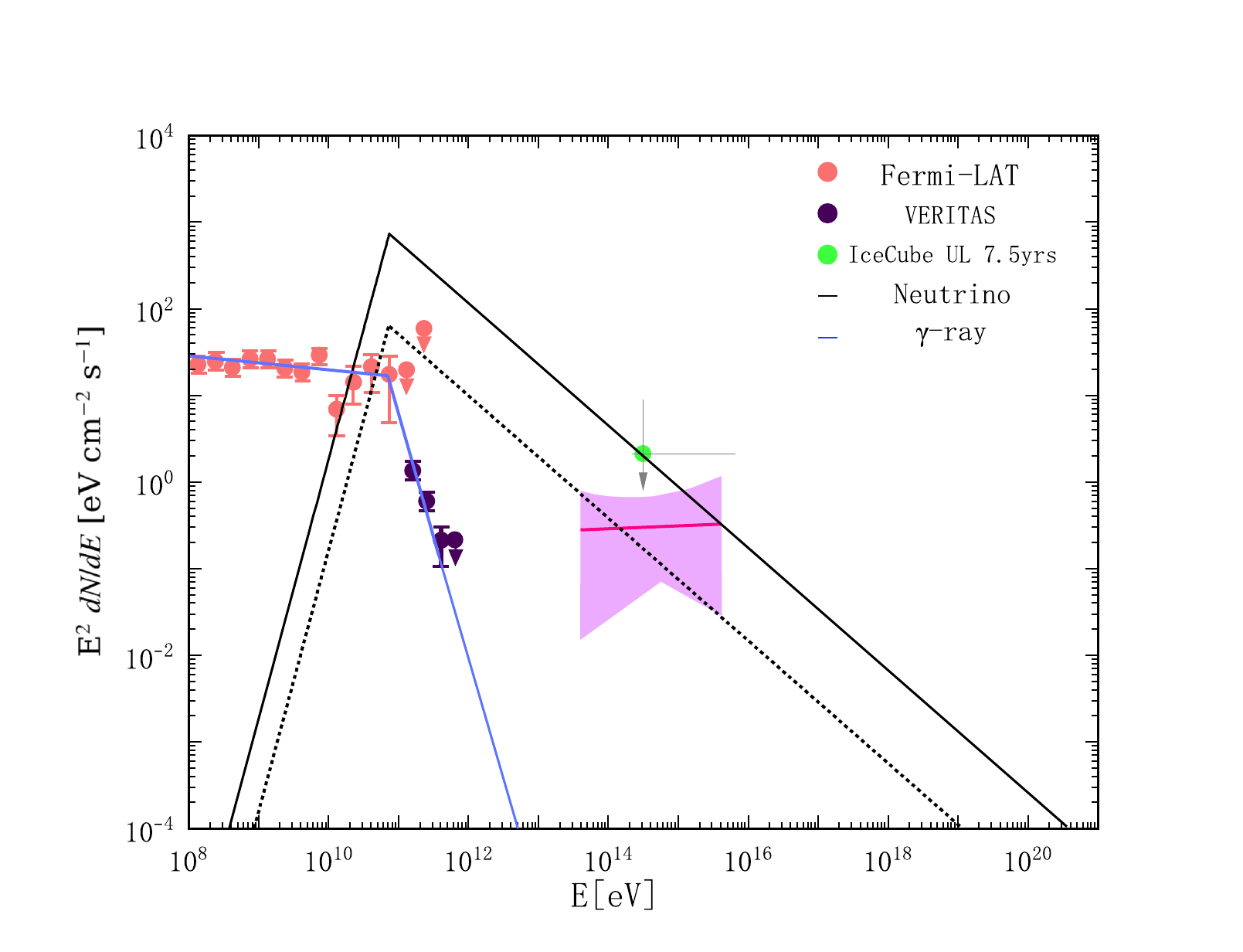}}
	\caption{The cosmic gamma-ray and neutrino spectra of TXS 0506+056 predicted by GC model. The blue solid line is the GC spectrum of   gamma-ray, the black solid and dashed lines correspond to the upper and lower limits of the predicted neutrino spectra, respectively. The  data of gamma-ray are from Fermi-LAT \cite{45} and VERITAS \cite{46}. The Green dot in the figure represents the 7.5-year upper limit on the neutrino flux reported by IceCube at an energy of $E_\nu = 311$ TeV \cite{3}, the magenta shaded region and solid line show the 10-year average neutrino flux inferred by \cite{4}. }
	\label{fig.1}
	
\end{figure}

The gamma-ray spectrum of TXS~0506+056, observed by Fermi-LAT \cite{45} and VERITAS \cite{46} over the period 2008--2023, exhibits a distinct broken power-law feature, as shown in figure~\ref{fig.1}. We interpret this feature as a signature of the GC effect in hadronic collisions. Within the GC model, the gamma-ray spectrum is described by eq.~(\ref{eq:2.9}) with parameters \(\beta_p = 1.85\), \(\beta_\gamma = 2.08\), \(C_\gamma = 2.3\times10^{-9}\), and a GC threshold \(E_\pi^{\mathrm{GC}} = 73.5\)~GeV (blue solid line in figure~\ref{fig.1}). 

The corresponding neutrino spectrum, given by eq.~(\ref{eq:2.10}), depends on the parameters $\beta_\nu$, $\beta_p$, and $C_\nu$. Here we adopt \(\beta_\nu = 0\), while $\beta_p$ is determined from the gamma-ray spectrum and $C_\nu$ must be constrained by experimentall data. Since the observations provide only an upper limit together with the range of the 10-year average flux, we determine $C_{\nu \max}$ from the upper limit and $C_{\nu \min}$ by requiring the spectral curve to pass through the observational range as closely as possible. These correspond to the solid and dashed black lines in figure~\ref{fig.1}, respectively. The resulting values are \(C_{\nu \max} = 1.55\times10^{-8}\) and \(C_{\nu \min} = 1.31\times10^{-9}\). We accounted for the attenuation of high-energy gamma rays by the extragalactic background light (EBL) using the \cite{47} model, and derived the intrinsic gamma-ray spectrum by applying the corresponding EBL correction, with the redshift fixed at z = 0.337 \cite{48}.

From figure~\ref{fig.1}, it can be seen that both the gamma-ray and neutrino spectra exhibit distinct GC features, with the characteristic GC break occurring around 73.5~GeV.

\subsection{Source NGC 1068}
\label{sec:3.2}

NGC 1068 is a prototypical Type II Seyfert galaxy located at a distance of 14.4~Mpc. It was first reported as a neutrino source with a significance of \(2.9\sigma\) in 2020, and subsequently confirmed at the \(4.2\sigma\) level in the 1–20~TeV energy range by IceCube 
 \cite{4,5}. The source emerged as the brightest hotspot in IceCube's 10-year all-sky survey, making it one of the most compelling neutrino-emitting AGN \cite{5}.

The gamma-ray spectrum of NGC~1068 is believed to consist of two distinct components: low-energy cascade emission from the AGN corona and high-energy emission from starburst activity. The starburst region, which is transparent to GeV--TeV gamma-rays, dominates the observed flux above \(\sim 100\)~MeV and is well described by hadronic \(pp\) interactions associated with star formation \cite{51,52}. In contrast, the corona around the central supermassive black hole is optically thick to high-energy gamma-rays because of the dense X-ray photon field, which reprocesses gamma-rays into lower energies through electromagnetic cascades \cite{40,41}. In this work, we consider only the gamma-rays and neutrinos produced by starburst activity.

\begin{figure}[htbp]
	\centering 
	{\includegraphics[width=1\textwidth]{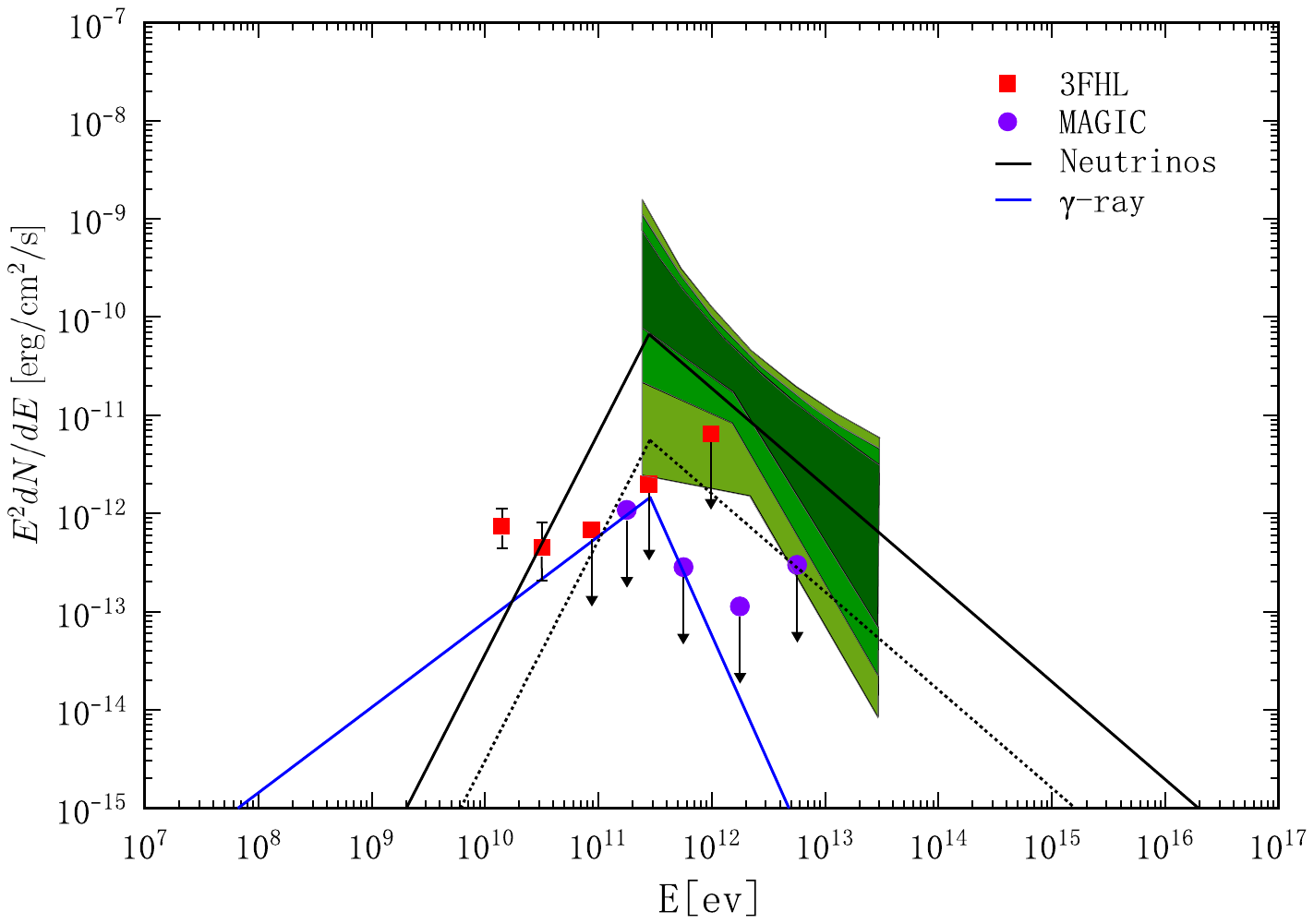}}
	\caption{The gamma-ray and neutrino energy spectra of NGC~1068 predicted by the GC model compared with the 3FHL \cite{53} and MAGIC \cite{54} data for NGC~1068. The green shaded regions represent the 1$\sigma$, 2$\sigma$, and 3$\sigma$ confidence regions of the neutrino spectrum measured by IceCube \cite{5}.The blue solid line is the GC spectrum of  gamma-ray, the black solid and dashed lines correspond to the upper and lower limits of the predicted neutrino spectra, respectively.}
	\label{fig.2}
\end{figure}

The gamma-ray spectrum of NGC~1068, as measured by 3FHL \cite{53} and MAGIC \cite{54}, exhibits a clear broken power-law feature, as shown in figure~\ref{fig.2}. We interpret this break as a signature of the GC effect in the starburst region. Using the GC model, we fit the gamma-ray spectrum with eq.~(\ref{eq:2.9}), adopting the parameters \(\beta_p = 2.00\), \(\beta_\gamma = 1.11\), and \(C_\gamma = 2.5\times10^{-12}\), with a GC threshold \(E_\pi^{\mathrm{GC}} = 280\)~GeV (blue solid line). 

The corresponding neutrino spectrum can be obtained from eq.~(\ref{eq:2.10}). Here, however, the parameter $C_\nu$ is not determined directly by fitting experimentall data; instead, it is inferred from the ratio relation between two different sources given in eq.~(\ref{eq:2.13}). Using TXS~0506+056 as the reference source and adopting its \(C_{\nu \max}\) and \(C_{\nu \min}\) values, we obtain \(C_{\nu \max} = 1.84\times10^{-11}\) and \(C_{\nu \min}= 1.54\times10^{-12}\) for NGC~1068. Using the \cite{47} model, the intrinsic gamma-ray spectrum was derived by applying the corresponding EBL correction, with the redshift fixed at z = 0.0037 \cite{55}.The corresponding spectra are shown as the black solid and dashed lines in figure~\ref{fig.2}. Compared with the experimentallly allowed energy range, our predictions lie within a reasonable interval.

Furthermore, relative to the experimentallly inferred neutrino spectral range, our predictions appear to be somewhat underestimated. This is most likely because we consider only the neutrinos originating from the starburst component and omit any contribution from the coronal component.

\subsection{Source SNR G54.1+0.3}
\label{sec:3.3}

In addition to the two AGN sources discussed above, the supernova remnant G54.1+0.3 has recently emerged as a candidate for coincident gamma-ray and neutrino emission. With an estimated age of \(\sim 2900\) years and a distance of \(\sim 6\) kpc, G54.1+0.3 is a young Crab-like SNR containing a pulsar wind nebula (PWN) powered by the 136-ms pulsar PSR J1930+1852 \cite{56,57,58,59,60,61}. The remnant exhibits a torus-like structure and jets similar to the Crab Nebula, and is considered a potential Galactic PeVatron capable of accelerating protons to PeV energies via diffusive shock acceleration \cite{6}.

Located approximately \(0.3^\circ\) west of the SNR is a molecular cloud known as the “Western Cloud,” with a mass of \(\sim 1.9\times10^4\,M_\odot\). The LHAASO observatory has detected an ultra-high-energy gamma-ray source, LHAASO J1929+1846u, whose position is spatially coincident with both the SNR and the Western Cloud. The KM2A array detected emission above 100~TeV that is concentrated in the Western Cloud region, suggesting that these gamma-rays originate from hadronic \(pp\) interactions between cosmic rays accelerated by the SNR and the ambient cloud material \cite{62}. Furthermore, a 67.6~TeV track-like neutrino event observed by IceCube \cite{63,64} is located within \(1.23^\circ\) of LHAASO J1929+1846u, raising the possibility that this neutrino shares a common hadronic origin with the ultra-high-energy gamma-rays.

\begin{figure}[htbp]
	\centering 
	{\includegraphics[width=0.9\textwidth]{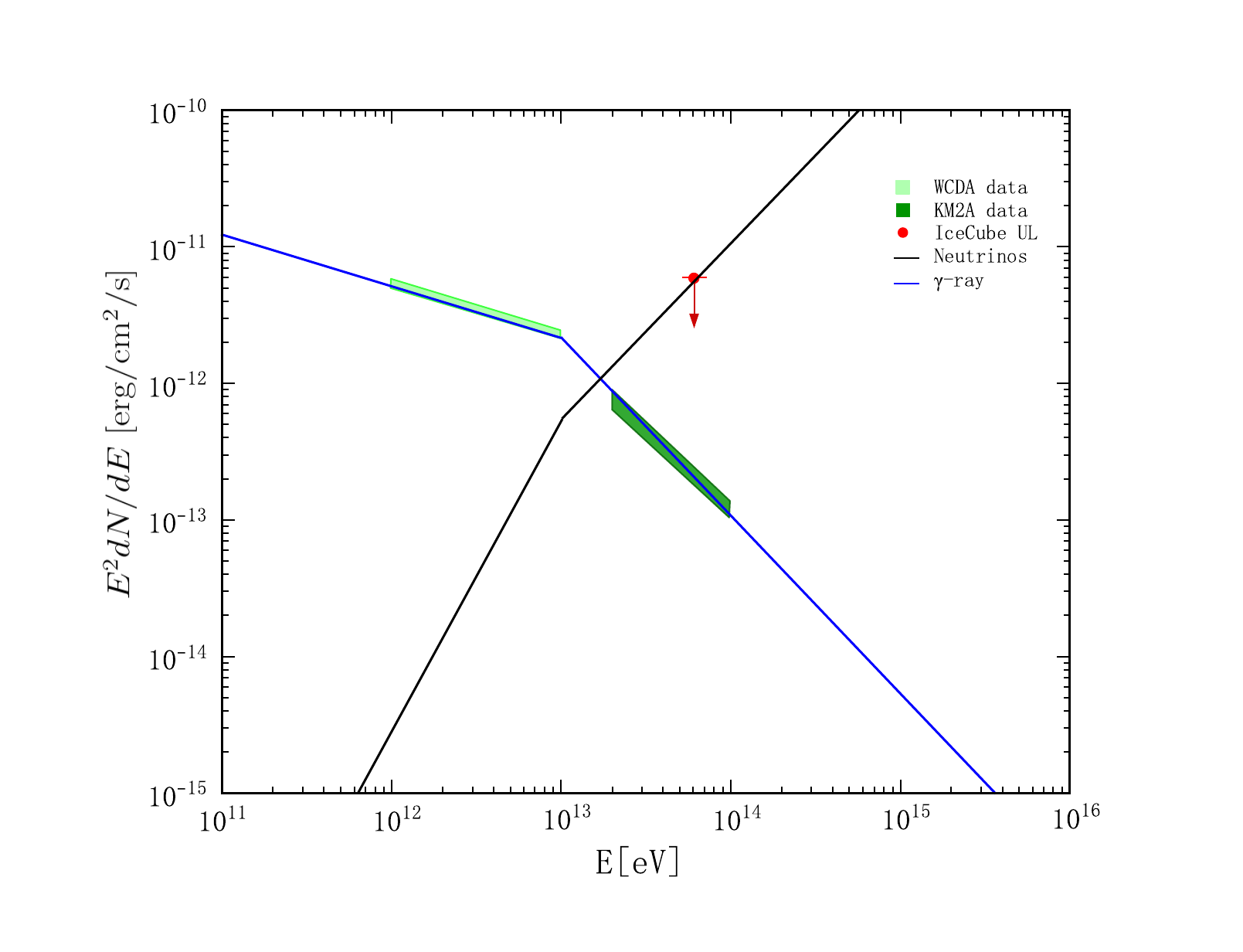}}
	\caption{The gamma-ray and neutrino spectra of SNR~G54.1+0.3 predicted by the GC model, together with data from WCDA and KM2A \cite{62}. The IceCube upper limits are shown as the green shaded region. The blue solid line is the GC prediction of  gamma-ray, and the black solid line is the corresponding prediction of neutrino spectra.}
	\label{fig.3}
\end{figure}

The gamma-ray spectrum of LHAASO~J1929+1846u, measured by the WCDA and KM2A arrays \cite{62}, exhibits a broken power-law feature with a break at approximately 10.06~TeV, as shown in figure~\ref{fig.3}. We nevertheless attempt to provide a unified description of both its gamma-ray and neutrino spectra within the framework of the GC model. For the gamma-ray spectrum, eq.~(\ref{eq:2.9}) provides a good fit with the parameters \(\beta_p = 0.95\), \(\beta_\gamma = 2.37\), \(C_\gamma = 2.49\times10^{-17}\), and a GC threshold \(E_\pi^{\mathrm{GC}} = 10.06\)~GeV (blue solid line). For the corresponding neutrino spectrum, using eq.~(\ref{eq:2.10}) together with the experimentall data, we adopt the parameters \(\beta_\nu = 0\) and \(C_\nu = 2.38\times10^{-18}\); the resulting spectrum is shown as the black solid line in figure~\ref{fig.3}.

As illustrated in the figure, the neutrino spectrum exhibits a continuous hardening trend above the GC break, with a high-energy spectral index of \(\sim 0.9\). This behavior deviates significantly from the typical power-law decay expected for cosmic-ray secondaries. Moreover, the parameter $C_\nu$ does not satisfy the relation with TXS~0506+056 and NGC~1068 implied by eq.~(\ref{eq:2.13}), which suggests that the neutrinos from SNR G54.1+0.3 and the measured gamma-ray spectrum most likely do not come from a GC process.                                                                 

Several factors may contribute to this discrepancy. First, the gamma-ray and neutrino emissions from this source may not share a common GC origin. The observed gamma-ray break could instead arise from other physical mechanisms, such as a cutoff in the parent proton spectrum or energy-dependent diffusion. Second, the special environment of the SNR--cloud interaction region---including shock effects, turbulent magnetic-field amplification, and non-equilibrium particle acceleration---could distort the spectral evolution relative to the ideal GC prediction. Third, the observed gamma-ray spectrum may be a superposition of multiple components, only some of which exhibit GC characteristics, thereby producing a composite spectral shape that does not conform to the single-source GC model. Therefore, not every gamma-ray spectrum with a broken power-law feature can be used for multi-messenger joint studies within the GC framework.

\section{Summary and discussion}
\label{sec:4}

In this work, we have performed a unified analysis of gamma-ray and neutrino spectra from astrophysical sources within the framework of the Gluon Condensation (GC) model. Starting from the common hadronic origin of secondary particles, we derived the neutrino spectrum corresponding to the GC gamma-ray spectrum and established an explicit relation between the two messengers. This makes it possible to use the observed gamma-ray spectrum to constrain the spectral parameters and then predict the associated neutrino emission in a self-consistent way.

We applied this framework to three representative sources with both gamma-ray observations and neutrino candidate signals: the blazar TXS~0506+056, the Seyfert galaxy NGC~1068, and the supernova remnant G54.1+0.3. For TXS~0506+056, the observed gamma-ray spectrum can be well fitted by the GC model, and the predicted neutrino spectrum is consistent with the IceCube upper limits and average flux range within uncertainties. For NGC~1068, the GC model also provides a good description of the gamma-ray spectrum, and the predicted neutrino flux, inferred through the relation between different sources, lies within a reasonable range compared with IceCube observations. These results suggest that TXS~0506+056 and NGC~1068 may share a common GC origin for their gamma-ray and neutrino emissions respectively.

By contrast, the case of SNR~G54.1+0.3 appears to be different. Although its gamma-ray spectrum also shows a broken power-law structure, the neutrino spectrum predicted by the GC model exhibits continuous hardening after the break. Such behavior deviates from the conventional expectation for cosmic-ray induced neutrino spectra and therefore disfavors a common GC interpretation for this source. This comparison indicates that not every broken gamma-ray spectrum should be directly attributed to the GC effect, and the neutrino counterpart provides an important additional diagnostic.

The main significance of this work is that it offers the first systematic attempt to connect gamma-ray and neutrino spectra under the GC framework. Compared with purely phenomenological descriptions, the GC model has stronger predictive power because both messengers are linked through the same underlying hadronic process. In particular, the relative magnitudes and spectral shapes of gamma-rays and neutrinos can be quantitatively tested against multi-messenger observations.

Nevertheless, several limitations should be noted. For example, some sources, especially AGN, multiple emission zones may contribute simultaneously, and the observed gamma-ray spectrum may be a superposition of different physical components; the current neutrino data still suffer from limited statistics, which restricts the precision of model tests,etc.

Future observations will provide a more stringent test of this scenario. With improved gamma-ray measurements from facilities such as NG-ACTA \cite{65} and more sensitive neutrino observations from IceCube-Gen2 \cite{66} and other next-generation detectors, the spectral relation predicted by the GC model can be examined more accurately. A broader application of this framework to additional candidate sources may also help determine whether the GC effect is a generic mechanism in high-energy astrophysical environments or only relevant for a subset of objects.

\acknowledgments

This work is supported by the National Natural Science Foundation of China (No. 11851303).



\bibliographystyle{JHEP} 
\bibliography{main} 



\end{document}